\documentclass[twocolumn,fleqn,,usenatbib]{mnras}
\usepackage{hyperref}
\usepackage{bm}
\usepackage{multirow}
\usepackage{gensymb}
\usepackage{enumitem}
\usepackage{tabularx}
\usepackage[flushleft]{threeparttable}
\usepackage[]{color}

\usepackage{float}
\usepackage{breqn}

\usepackage{amsfonts,amsmath,amssymb,esint}
\usepackage{graphicx}
\usepackage{color}
\usepackage{xcolor}
\usepackage{url}

\title[Continuum Driven Stellar Instabilities]{Line Driven Instabilities due to Continuum Radiation Transport in Stellar Winds}

\author[S. Dyda]{
Sergei Dyda,$^{1}$\thanks{sergei@virginia.edu}
Shane W. Davis,$^{1}$
\\
$^{1}$ Department of Astronomy, University of Virginia, 530 McCormick Rd., Charlottesville, VA 22904, USA
}

\begin{document}

\label{firstpage}
\pagerange{\pageref{firstpage}--\pageref{lastpage}}

\maketitle

%%%%%%%%%%%%%%%%%%%%%%%%%%%%%%%%%%%%%%%%%%
\begin{abstract}
We study line driven stellar winds using time-dependent radiation hydrodynamics where the continuum radiation couples to the gas via either a scattering or absorption opacity and there is an additional radiation force due to spectral lines that we model in the Sobolev approximation. We find that in winds with scattering opacities, instabilties tend to be suppressed and the wind reaches a steady state.  Winds with absorption opacities are unstable and remain clumpy at late times. Clumps persist because they are continually regenerated in the subcritical part of the flow. Azimuthal gradients in the radial velocity distribution cause a drop in the \emph{radial} radiation force and provide a mechanism for generating clumps. These clumps form on super-Sobolev scales, but at late times become Sobolev-length sized indicating that our radiation transfer model is breaking down. Inferring the clump distribution at late times therefore requires radiation-hydrodynamic modeling below the Sobolev scale.   
\end{abstract}

%%%%%%%%%%%%%%%%%%%%%%%%%%%%%%%%%%%%%%%%%%%%%%%%%%%%%%%%%%%%%%%%%%%%%
\section{Introduction}
\label{sec:introduction}

Radiation pressure due to spectral lines is thought to provide the force to drive outflows from a variety of compact object systems, such as OB stars, cataclysmic variables (CVs) and active galactic nuclei (AGN). The possibility of ejecting metalic ions from stars via radiation pressure on spectral lines was recognized by \cite{Milne1926}. Subsequently, \cite{Lucy1970} and \cite{CAK1975}, (hereafter CAK) showed that if these metal ions were well coupled to $H$ and $He$ gas this mechanism could drive significant winds from massive stars.

Correctly modeling line driven outflows has required progress in two important areas. Firstly, photoionization studies are needed to predict the number and strength of lines, which ultimately determines the coupling strength between the radiation and the gas i.e the force multiplier. Secondly, one must understand the radiation transfer of the irradiating flux through the wind. These problems are intimately coupled i.e the radiation transfer depends on the state of the gas and the state of the gas depends on the irradiating SED. Due to computational limitations, efforts have been made to attack these coupled problems in approximate ways.

Photoionization studies have used atomic line lists to compute the strengths of radiative driving. \cite{Abbott1982} showed that for type O-G stars the line force was relatively constant. \cite{Gayley1995} extended the original CAK formalism to include strong line overlap effects using a non-isotropic diffusion approximation. \cite{Pauldrach1994} accounted for some non-LTE effects in photoionization modelling of the most important lines in the context of O stars. \cite{Owocki1996} improved on the line force in CAK by accounting for integral escape probabilities in the radiation transfer of line photons.  \cite{Puls2000} further refined this approach to improve the calculation of radiation force from line statistics.     

The earliest studies made use of the Sobolev approximation, whereby the wind was assumed to be optically thin to the continuum but could extract momentum from the radiation field at resonance points in the wind where the Doppler shift was equal to the frequency difference between the continuum photons and the line. The strength of the Sobolev approximation is that one may express the line force as a function of only \emph{local} variables, rather than have to resolve the radiation transfer of a line through the entire flow, which up until now has been computationally intractable. In the original  CAK model the strength of the radiation force depends only on the radial velocity gradients $dv/dr$ through the optical depth parameter
\begin{equation}
    t = \frac{\rho v_{\rm{th}} \sigma_e}{dv/dr},
\label{eq:opticaldepth}    
\end{equation}
where $\rho$ is the density, $ v_{\rm{th}}$ the thermal velocity and $\sigma_e$ is the mass scattering coefficient for free electrons.   Later models extended this formalism to include azimuthal velocity gradients \citep{Grinin1978} which was applied to study corotating interaction regions (CIRs) (\cite{Cranmer1995}; \cite{Dessart2004}) and rotating hot stars \citep{Gayley2000}.   

In the context of massive stars, models of radiative driving successfully reproduce typically observed outflow velocities and mass loss rates (for a review see \cite{Puls2008}). However, reconciling the observed mass flux with spectral modelling has relied on so called \emph{micro clumping}, density structures on scales smaller that the grid resolution of simulated outflows. Mass loss rates are typically inferred from emission lines, with an amplitude that scales $\sim \rho^2$ and is highly sensitive to inhomogeneities in the flow. Understanding the formation and evolution of density inhomogeneities or clumps is thus crucial for studying these systems.

The earliest studies recognized that line driven outflows are inherently unstable. \cite{Lucy1984} showed that scattering due to spectral lines would lead to a ``line drag'' instability. Likewise \cite{Owocki1988} showed that shadowing of gas in the outer parts of the flow by density perturbations would grow, the so called line deshadowing instability (LDI). \cite{Abbott1980} showed that accoustic waves propagating in and out from the critical point could alter the resulting wind solutions. This work, along with the observation that winds should be clumpy encouraged a lot of work into the generations of small scale structures in line driven winds. 

\cite{MacGregor1979} showed that waves are amplified in the optically thick parts of the flow. \cite{Martens1979} showed that sound waves are amplified by velocity gradients. \cite{Carlberg1980} found instabilities due to line shape and radiation and showed that driven sound waves will grow and change the wind dynamics. \cite{Owocki1984} showed that absorption lines lead to amplification of perturbations on time scales short compared to dynamical time. In a series of follow up papers, they subsequently showed that scattering effects, the so called ``line drag'' effect reduces these instabilities at the base of the wind and that disturbances propagate but do not grow \citep{Owocki1985} and perturbations propagate at the speed of sound \citep{Owocki1986}. They extended their analysis to include the finite disc effect and line dragging in multiple dimensions and showed the finite disc causes unstable lateral waves but these are subdominant to radial waves \cite{Rybicki1990}. \cite{Owocki1988} showed that in a pure absorption model, perturbations form at the base of the wind, leading to steep rarefied wave and shocks. 

In this paper we explore the formation of clumps due to time-dependent evolution of the continuum radiation field driving a stellar outflow. We extend the optically thin formulation of the classical line driving prescription to one where we resolve the time-dependent radiation transfer of continuum photons. Thus, though we still rely on a \emph{local} Sobolev approximation for modeling the line transfer, we account for both scattering and absorption effects of the continuum through the flow.

Continuum absorption tends to be unstable in the flow. The optical depth parameter (\ref{eq:opticaldepth}) will be larger for over-densities, hence the driving force $F_{\rm{rad}} \sim k t^{-\alpha}$ will tend to be smaller and the over-density will hence slow and tend to grow. By contrast, continuum scattering tends to isotropize the radiation field, so any density dependent effect will tend to be washed out.

We perform a series of simulations exploring the relative effects of scattering and absorption on 2D spherically symmetric, line driven winds. We explore how initial density perturbations tend to grow in the case of continuum absorption and are washed out in the case of continuum scattering. We characterize the growth of density and velocity fluctuations in terms of clumping parameters and use these to quantify the growth of the instabilities. 

The structure of this paper is as follows. In Sec \ref{sec:theory} we introduce our numerical methods, in particular our treatment of the radiation force. In Sec \ref{sec:results} we describe our results, namely the growth and saturation of density and velocity perturbations and characterize the clumpiness of the flow. Finally in Sec \ref{sec:discussion} we discuss the significance of these results for stellar winds. We further discuss how these simulations set-up the groundwork for our later study of single line transfer using our time-dependent radiation transfer method.

\section{Simulation Setup}
\label{sec:theory}

\subsection{Basic Equations}
\label{sec:equations}

The basic equations for single fluid radiation hydrodynamics are
\begin{subequations}
\begin{equation}
\frac{\partial \rho}{\partial t} + \nabla \cdot \left( \rho \mathbf{v} \right) = 0,
\end{equation}
\begin{equation}
\frac{\partial (\rho \mathbf{v})}{\partial t} + \nabla \cdot \left(\rho \mathbf{vv} + \sf{P} \right) =  \mathbf{G} + \rho \mathbf{g}_{\rm{grav}},
\label{eq:momentum}
\end{equation}
\begin{equation}
\frac{\partial E}{\partial t} + \nabla \cdot \left( (E + P)\mathbf{v} \right) = cG^{0} + \rho \mathbf{v} \cdot \mathbf{g}_{\rm{grav}},
\label{eq:energy}
\end{equation}
\label{eq:hydro}%
\end{subequations}
where $\rho$ is the fluid density, $\mathbf{v}$ the velocity, $\sf{P}$ a diagonal tensor with components $P$ the gas pressure. The total gas energy is $E = \frac{1}{2} \rho |\mathbf{v}|^2 + \mathcal{E}$ where $\mathcal{E} =  P/(\gamma -1)$ is the internal energy and $\gamma$ the gas constant. The gravitational source is due to a star with
\begin{equation}
    \mathbf{g}_{\rm{grav}} = -\frac{GM}{r^2} \hat{r}, 
\end{equation}
where $M$ is the stellar mass and $G$ the gravitational constant. The temperature is $T = (\gamma -1)\mathcal{E}\mu m_{\rm{p}}/\rho k_{\rm{b}}$ where $\mu$ is the mean molecular weight and other symbols have their standard meaning.  

The radiation source terms $\mathbf{G}$ and $cG^0$ are assumed to receive contributions from the continuum and spectral lines
\begin{subequations}
\begin{equation}
\mathbf{G} = \mathbf{G}_{\rm{cont.}} + \mathbf{G}_{\rm{lines}}, 
\end{equation}
\begin{equation}
G^{0} = G^{0}_{\rm{cont.}} + G^{0}_{\rm{lines}}. 
\end{equation}    
\end{subequations}

The continuum radiation field is treated by directly solving the time dependent radiation transport equation using the implicit implementation in \textsc{Athena++} \citep{Jiang2021}.  We refer the reader to \cite{Jiang2021} for the specific formulation but the equation solved is equivalent to
\begin{dmath}
\frac{\partial I}{\partial t} + c \mathbf{n} \cdot \nabla I = c S_I, 
\label{eq:dIdt}
\end{dmath}
with the source term
\begin{equation}
    S_I = \Gamma^{-3} \rho \left[ \left( \kappa_P \frac{c a T^4}{4\pi} - \kappa_{E} J_0 \right) - \left( \kappa_s + \kappa_{F} \right) \left(I_0 - J_0 \right) \right],
    \label{eq:source}
\end{equation}
where $\kappa_s$ is the scattering opacity, $\kappa_{F}$ is the absorption contribution to the flux mean opacity, $\kappa_P$ the Planck mean and $\kappa_E$ the energy mean opacity. $I_0$ is the intensity in the comoving frame and 
\begin{equation}
    J_0 = \frac{1}{4\pi}\int I_0 \; d\Omega_0,
\end{equation}
is the corresponding angle averaged comoving frame mean intensity.

We consider a restricted class of models where we either set $\kappa_s = \kappa_{\rm es}$ (scattering model, ``Model S"), or $\kappa_{F} = \kappa_{E} = \kappa_{\rm es}$ (absorption model, ``Model A"). In both cases, all other opacities are set to zero. Here, $\kappa_{\rm es}$ is the electron scattering opacity, which is approximated as constant in the Thomson limit. 

The source term (\ref{eq:source}) then becomes
\begin{dmath}
S_I = 
\begin{cases} 
      -\Gamma^{-3} \rho \kappa_{es}(I_0 - J_0)  & \rm{scattering} \\
            -\Gamma^{-3} \rho  \kappa_{es}I_0  & \rm{absorption} 
   \end{cases}
\label{eq:dIdt}
\end{dmath}
The absorption case thus corresponds to a pure attenuation case where the radiation field is simply attenuated. In the scattering case, the mean intensity acts as a local, isotropic (in the co-moving frame) radiation source.

With the above assumptions the continuum momentum and energy source terms are then
\begin{equation}
    \mathbf{G}_{\rm{cont.}} = \frac{1}{c}\int \mathbf{n} S_{I} d \Omega,
\end{equation}
\begin{equation}
    cG^0_{\rm{cont.}} = c \int S_{I} d \Omega.
\end{equation}

We model the force due to lines via a CAK type prescription using the local continuum flux. Working in the Sobolev approximation, the line force
\begin{equation}
\mathbf{G_{\rm{lines}}} = \frac{\rho \kappa_{es}}{c} \oiint M(t) \mathbf{n} I(\mathbf{n}) d\Omega,
\end{equation}
where the integral is over all radiation rays of the continuum. The strength of the spectral lines, relative to electron scattering is quantified via the
CAK prescription
\begin{equation}
    M(t) = k t^{-\alpha},
\end{equation}
where $k = 0.2$ and $\alpha = 0.6$ is the ratio of optically thick to optically thin lines. The optical depth parameter 
\begin{equation}
    t = \frac{\rho v_{\rm{th}} \sigma_e}{|dv/dl|},
    \label{eq:dvdl}
\end{equation} 
where $v_{\rm{th}}$ is the gas thermal velocity and $dv/dl$ the velocity gradient along the line of sight of the radiation flux.
In general, this gradient can be expressed as
\begin{equation}
    \frac{dv}{dl} = \varepsilon_{ij}n_i n_j,
    \label{eq:epsilon_ij}
\end{equation}
where $\mathbf{n}$ is the unit vector in the direction of the incident radiation and the components of the shear tensor
\begin{align}
\varepsilon_{ij} n_i n_j &= 
\frac{\partial v_r}{\partial r} n_r n_r 
+ \frac{1}{r} \left( \frac{\partial v_{\theta}}{\partial \theta}  + v_r \right) n_{\theta} n_{\theta} \\ 
&+ \frac{1}{r \sin \theta} \left( \frac{\partial v_{\phi}}{\partial \phi} + v_r \sin \theta + v_{\theta} \cos \theta \right) n_{\phi} n_{\phi} \\ 
&+ \left( \frac{1}{r} \frac{\partial v_r}{\partial \theta} + \frac{\partial v_{\theta}}{\partial r} - \frac{v_{\theta}}{ r}\right) n_r n_{\theta} \\
&+ \left( \frac{1}{r \sin \theta} \frac{\partial v_{r}}{\partial \phi} + \frac{\partial v_{\phi}}{\partial r} - \frac{v_{\phi}}{r} \theta\right)  n_{r} n_{\phi} \\
&+ \frac{1}{r}\left( \frac{1}{\sin \theta} \frac{\partial v_{\theta}}{\partial \phi} + \frac{\partial v_{\phi}}{\partial \theta} - v_{\phi}\cot \theta\right)  n_{\theta} n_{\phi}.
\end{align}
 In the CAK case, radiation is assumed to be from a point source and only the $rr$ component contributes so we have
\begin{equation}
\frac{dv}{dl} \approx \frac{dv_r}{dr}.
\label{eq:dvdr}
\end{equation}
This is sometimes referred to as as the radial streaming approximation. 

More sophisticated prescriptions of the force multiplier have been developed (see for example \cite{Owocki1988}), but we will use the CAK prescription for ease in comparing to analytic results. 

The work done by the line force is then
\begin{equation}
G^0_{\rm{lines}}(E) = v \cdot \mathbf{G_{\rm{lines}}}. 
\end{equation}

To better understand the effects of scattering and absorption of continuum radiation we will compare our results to the original CAK formulation, where the wind is assumed to be optically thin to the continuum and the force due to continuum and lines is 
\begin{equation}
\mathbf{G_{\rm{CAK}}} = \Gamma \Big(M(t) + 1\Big)\frac{GM}{r^2} \hat{r}, 
\end{equation}
where $\Gamma = L_{*} \sigma_e / 4\pi cGM$  is the Eddington fraction with $L_*$ the stellar luminosity and $M(t)$ is the force multiplier. The corresponding energy source term is then
\begin{equation}
    G^0_{\rm{CAK}} = v \cdot \mathbf{G_{\rm{CAK}}}. 
\end{equation}

\subsection{Simulation Parameters}
We choose simulations relevant to an O star. We take the mass to be $M = 50 M_{\odot}$ and radius $r_* = 10 r_{\odot}$. We set the Eddington fraction $\Gamma = 0.1$ and the line driving parameters $k = 0.2$ and $\alpha = 0.6$. The temperature of $T = 10^{5}$ K  ensures that thermal driving is negligible \citep{Stone2009}. We use a unit of time $t_0 = 10^2$ s, which is a typical timescale for perturbations to propagate in the flow.  

The simulation region extends radially from $r_* < r < 10 r_*$ and we take a wedge $\Delta \theta = 0.2$ with $N_{\theta} = 64$. This dynamical range allows the wind to fully accelerate and reach its terminal velocity. We use a logarithmically spaced grid of $N_r$ = 1024 points and a scale factor $a_r = 1.008$ that defines the grid spacing recursively via $dr_{n+1} = a_r dr_{n}$. This resolution ensures that we resolve up to the Sobolev length, where our model for the line transfer breaks down.  

At the inner boundary, we impose outflow boundary conditions on $v$ and $E$ while keeping the density fixed at $\rho_* = 10^{-10}$ $g/cm^3$ in the first active zone. This density ensures that we are sufficiently resolving the atmosphere at the base of the wind, but also providing sufficient mass to launch a wind given our choice of Eddington parameter. The continuum intensity of the radially outgoing rays is set by the Eddington parameter while the non-radial rays are set to zero.

At the outer boundary, we impose outflow boundary conditions on $\rho$, $v$ and $E$ and vacuum conditions on the continuum. We use periodic boundaries along the azimuthal boundaries.

\section{Results}
\label{sec:results}
We study two line driven wind models. 1) An \emph{absorption} model, where $\kappa_E = \kappa_F =\kappa_{es}$ which we will refer to as Model A. 2) a \emph{scattering} model where $\kappa_s = \kappa_{es}$, which we will refer to as Model S. All other opacities are set to zero. 

\subsection{Bulk Flow}

\begin{figure*}
    \centering
    \includegraphics[scale=0.42]{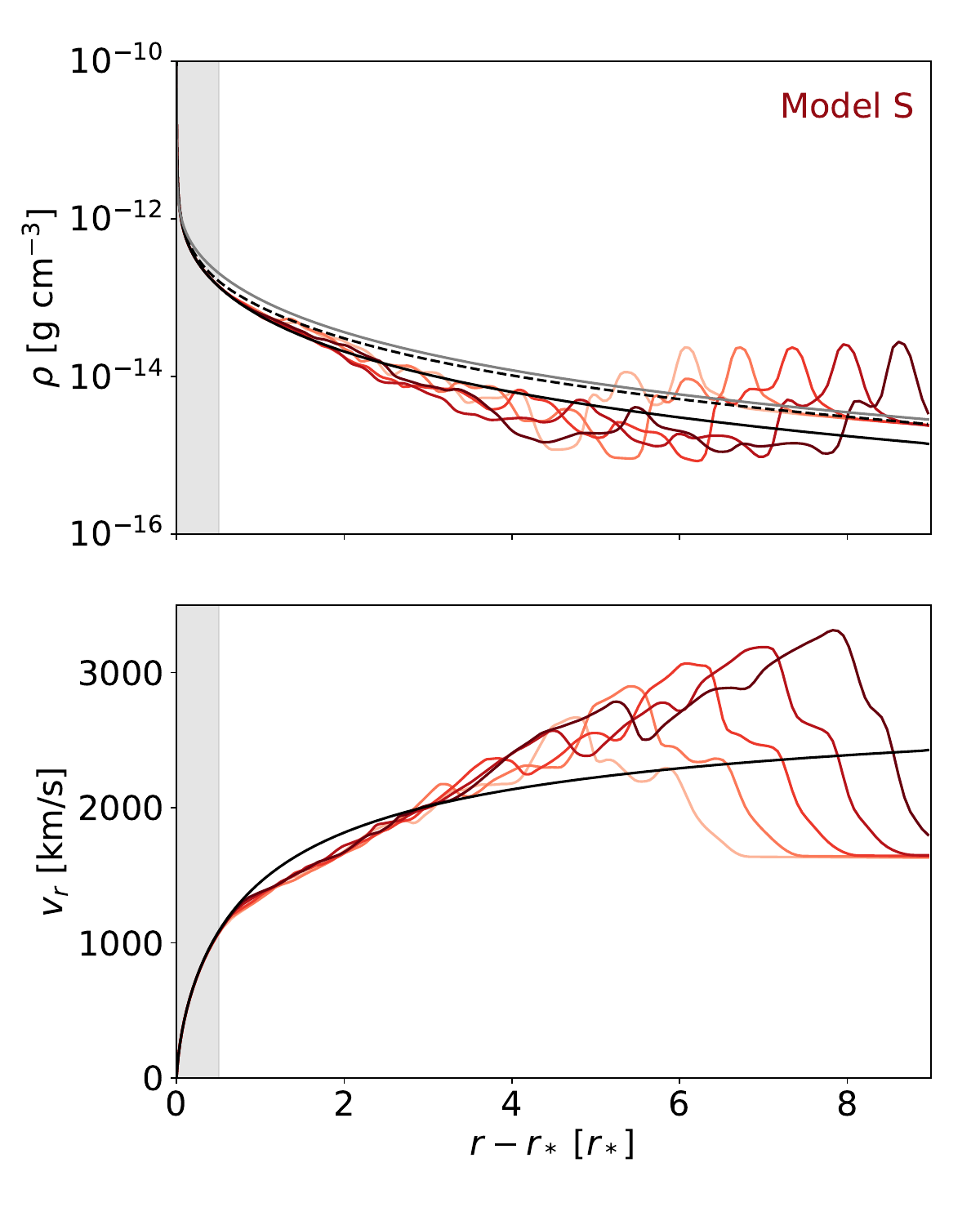}
    \includegraphics[scale=0.42]{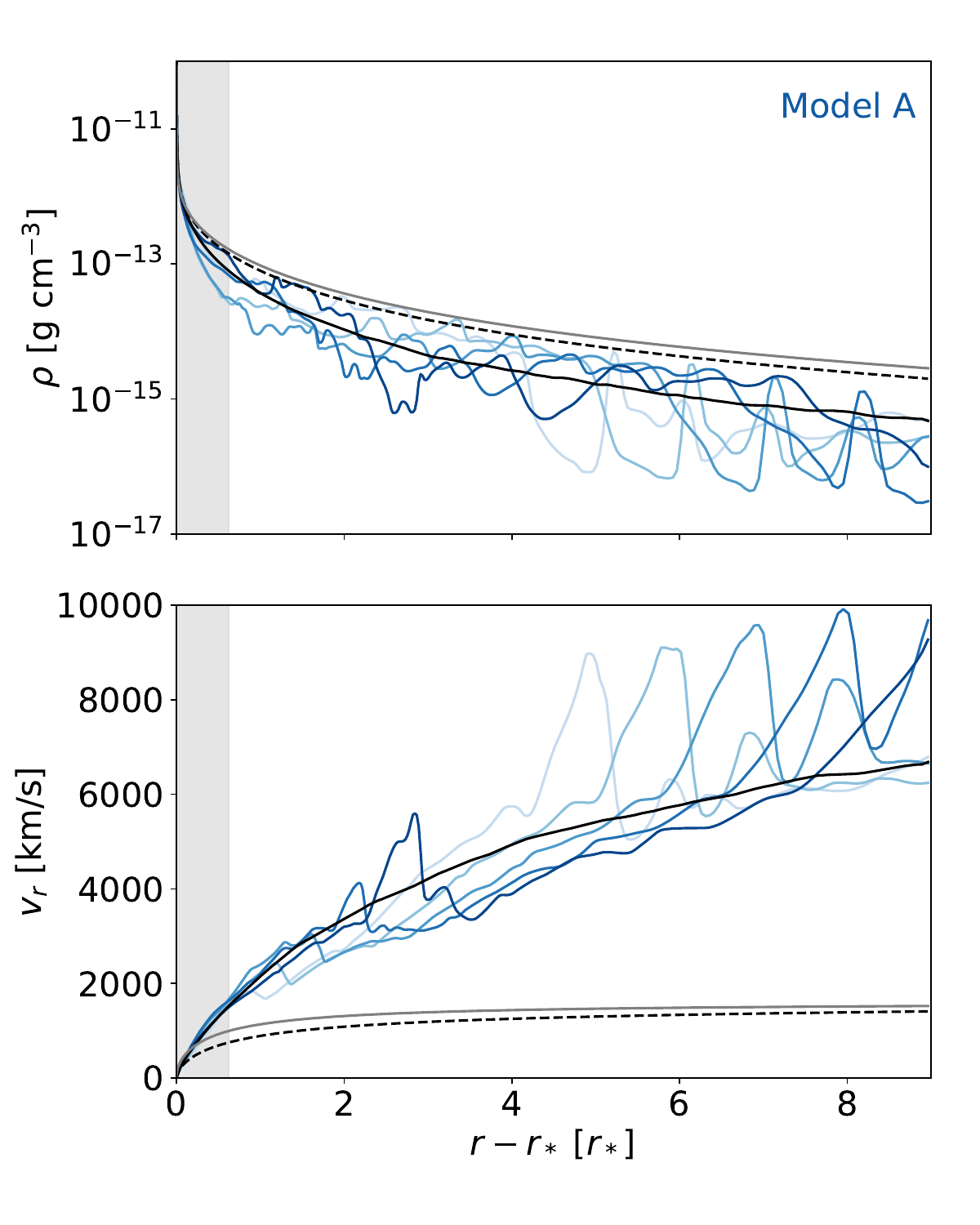}
    \caption{Density (top panel) and velocity (bottom panel) for Model S (left side panels) and Model A (right side panels). We show the time-averaged solution using the radial streaming approximation (dashed black line) and the full expression for the velocity gradient (\ref{eq:epsilon_ij}) (solid black line), and 5 progressively later (darker curves) snapshots in time with $\Delta t = 10 t_0$. We plot the CAK solution in grey, and indicate the sub-critical region of the model with the grey shading.}
    \label{fig:perturbations}
\end{figure*}

\begin{figure}
    \centering
    \includegraphics[scale=0.4]{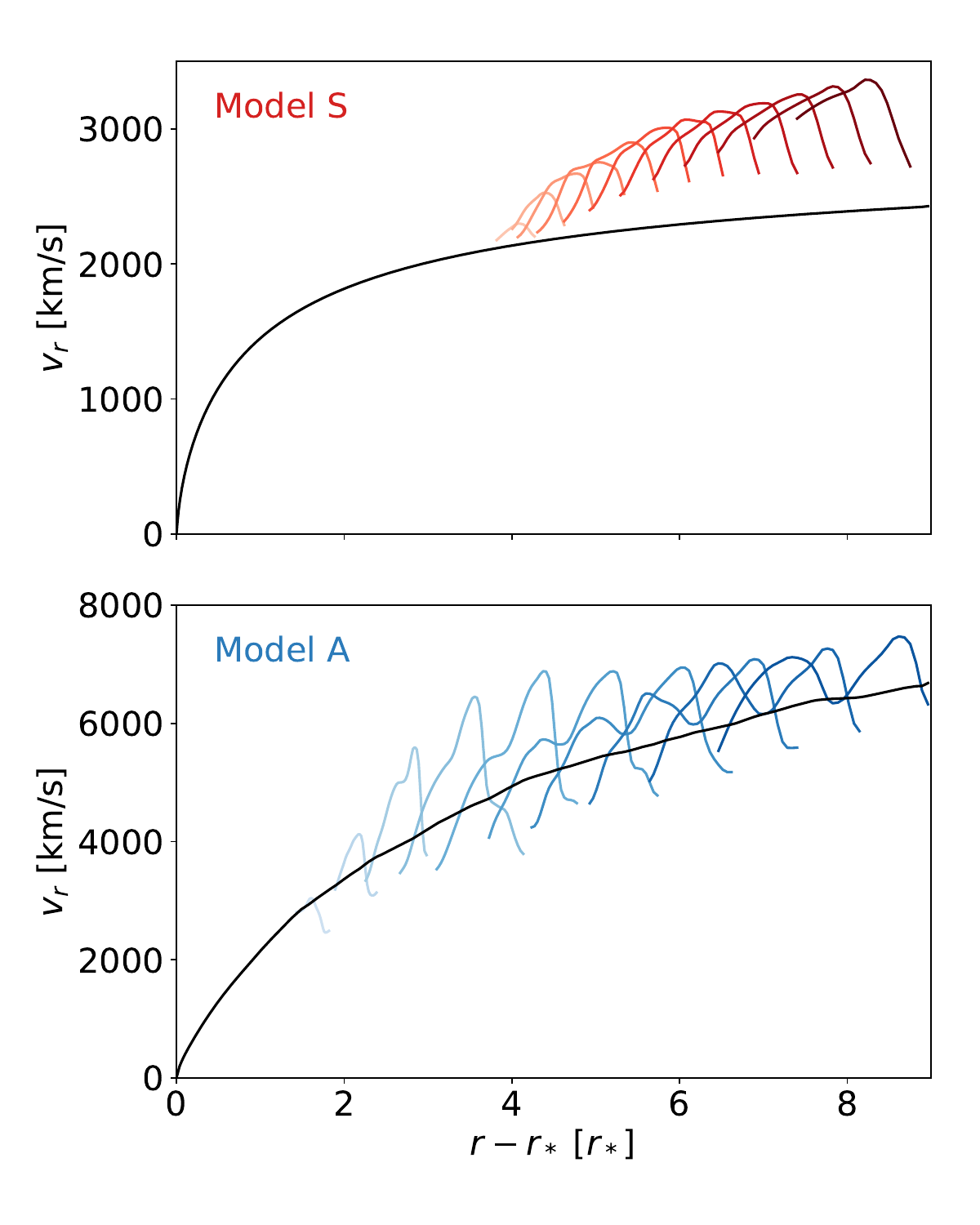}
    \caption{Fixed time snapshots of a feature in the velocity profile at time intervals $\Delta t = 10 t_0$, where lighter shades are from earlier times. We observe the amplitude of the wave increasing and the waveform diffusing, with the diffusion effect more prominent in Model A.}
    \label{fig:wave-steepening}
\end{figure}

\begin{figure}
    \centering
    \includegraphics[scale=0.4]{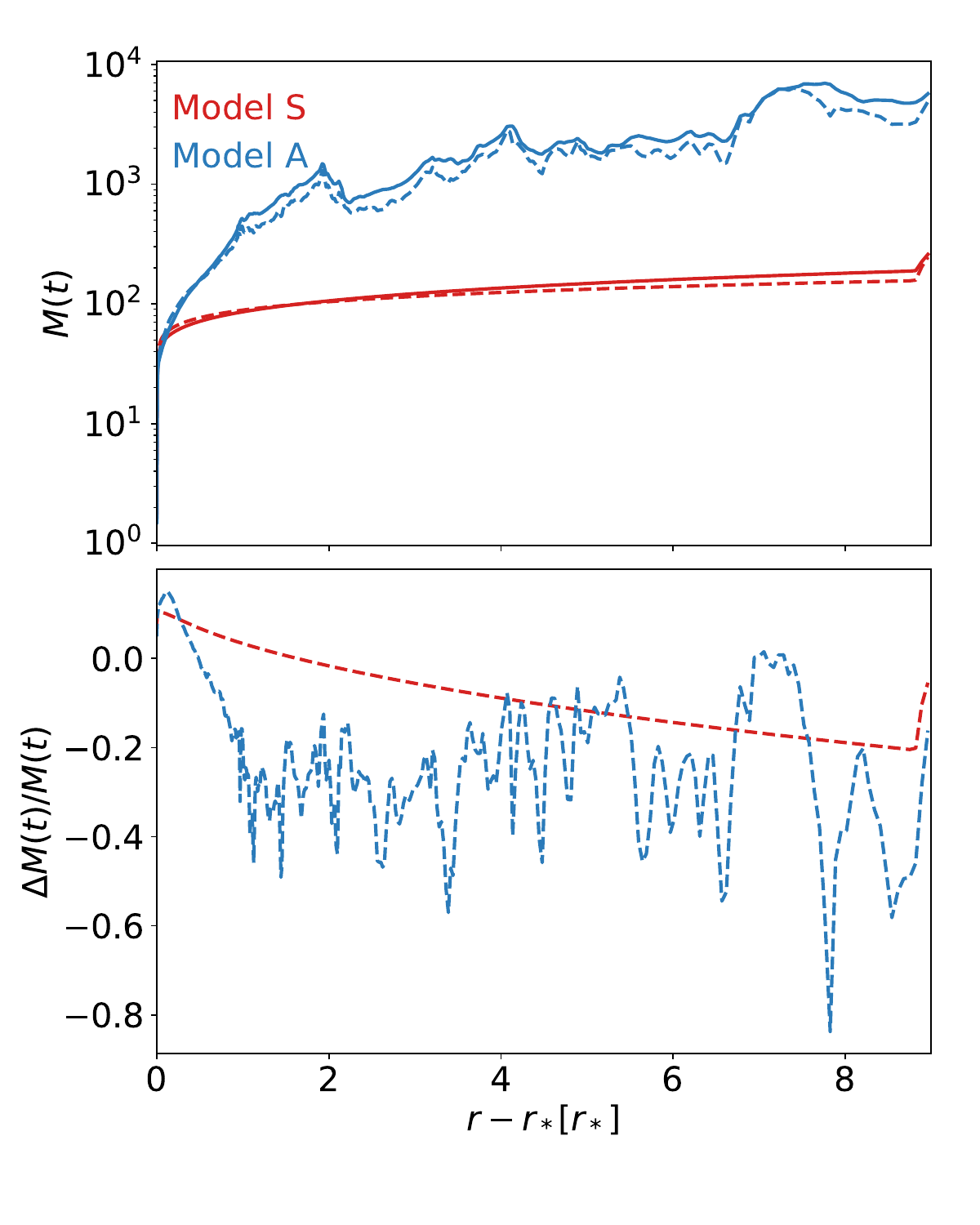}
    \caption{\textit{Top panel -}Time-averaged force multiplier at late times for Model S (red lines) and Model A (blue lines) using time dependent radiation transport (solid lines). For comparison, we show the CAK optically thin force multiplier (dashed lines). \textit{Bottom panel} Difference in force multipliers, normalized to the optically thin value.}
    \label{fig:Mt_avg}
\end{figure}

In Fig. \ref{fig:perturbations} we plot the density and velocity profiles for Models S and A. The solid black lines show the time-averaged late time behaviour and the dashed black line shows the analytic CAK solution. The grey line indicates the solution using the radial streaming approximation to compute the velocity gradient in the force multiplier. The colored lines indicate snapshots separated in time by $\Delta t = 10 t_0$, with the darker lines being later in time. The grey shading indicates the subcritical region of CAK, where the critical point $r_c$ is defined by
\begin{equation}
    \frac{dv}{dr}\Big|_{r_c} = \frac{v(r_c)}{r_c}.
\end{equation}

The models differ crucially in terms of their late time behaviour. Model S reaches a steady state, relatively close to the CAK solution, whereas Model A is unstable, with a terminal velocity $\sim 3 $ times greater than CAK.

In Model S, perturbations continue to grow until $t = 700 \ t_0$, roughly corresponding to the dynamical time of the wind. This is the epoch shown in these snapshots. At late times, the perturbations die out and the solution settles to the solid black line shown. Conversely, in Model A perturbations are long lived and the flow never reaches a steady state. The late-time averaged solution is smooth but perturbations are continuously created in the sub-critical region and advected across the flow. We have verified that these instabilities persist by running simulations up to $t = 9000 \ t_0$ and also increasing the domain size to $r_* \leq r \leq 20 r_*$.

We note that density and velocity perturbations are out of phase by $\Delta \phi = \pi$, as predicted by the linear analysis for a pure absorption model in  \citep{Owocki1988}.

From this series of snapshots we can infer the speed of the outgoing waves. In Fig. \ref{fig:wave-steepening} we plot a wave feature in the velocity profile at fixed $\theta$ for time intervals $\Delta t = 10 \ t_0$, with darker shades being at later times for both models.  We observe the amplitude of the wave increasing, and the waveform diffusing. The diffusion effect is stronger for Model A. We have plotted the initial feature at the earliest time it became visible. Consistent with our previous observations of the density and velocity profiles, the perturbations in Model S form at larger radii than in Model A. 

The difference in the two models is not due to instantaneous differences in the force multiplier. In Fig \ref{fig:Mt_avg} we plot the time-averaged force multiplier for both models. The solid lines are for the time-dependent radiation transfer and the dashed lines are for the optically thin treatment. In the bottom panel we plot the relative difference between the two. We notice that the difference is in the $\sim 10\%$ and $\sim 30\%$ range for models S and A respectively. This suggests that the divergence in the two solutions is not due to the instantaneous radiation transfer. In fact, for a given fluid distribution the optically thin and optically thick treatments yield similar values of the force multiplier. We do not expect exact agreement anyways since the optically thin treatment uses the radial streaming approximation for the velocity gradient. Therefore, the divergence in the two solutions is an aggregate effect over time, whereby clumps in Model A allow for an enhanced radiation force $M(t) \sim 10^3$ rather than $M(t) \sim 10^2$ for the scattering model. Here we have not assumed a maximum value of the force multiplier as might be expected to occur due to line saturation. Our models only serves to illustrate that that the line force may become large due to clumping.

\subsection{Clumpiness}

\begin{figure*}
    \centering
    \includegraphics[scale=0.75]{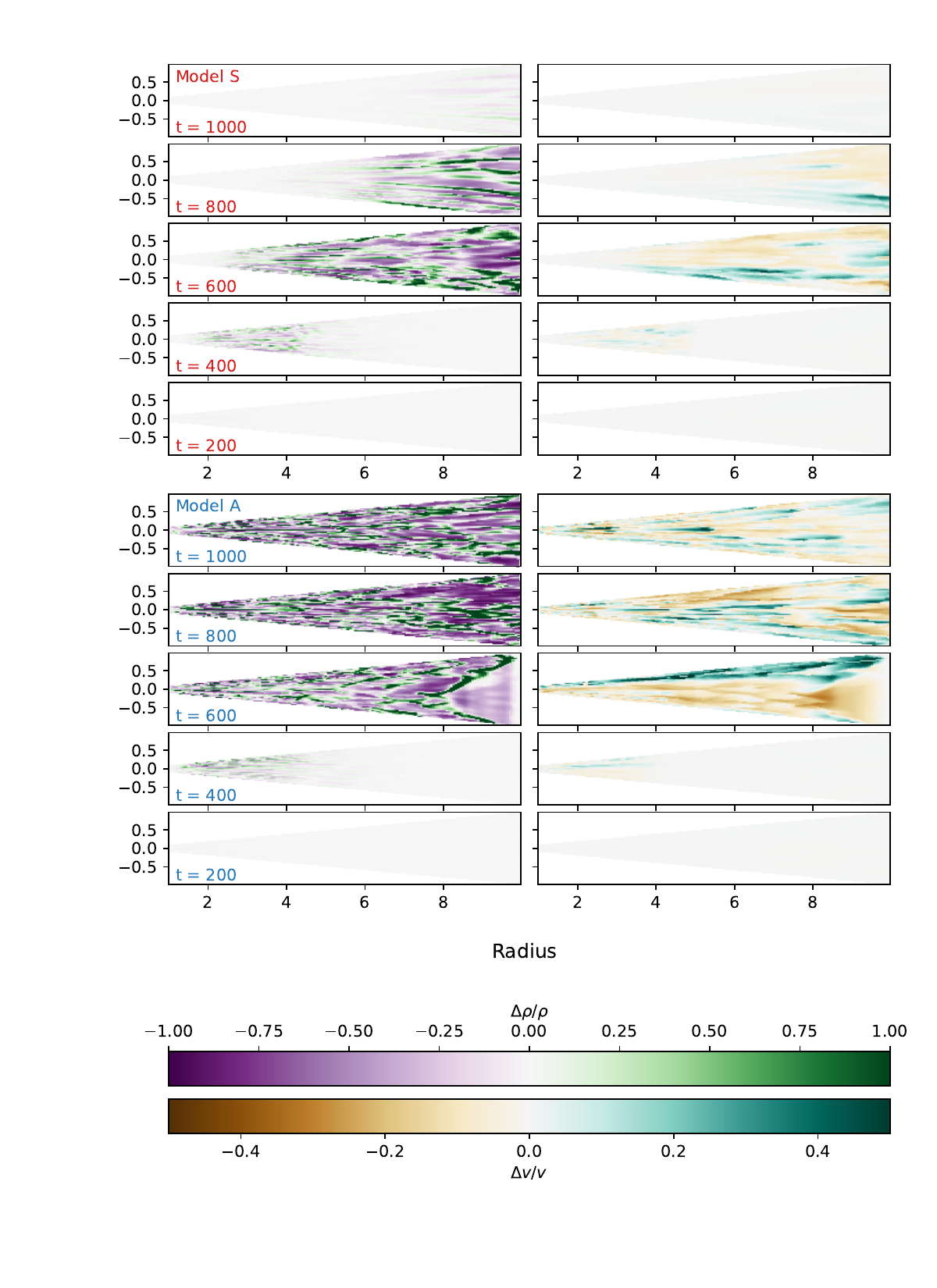}
    \caption{\textit{Top panels} - Model S density (left panels) and velocity (right panels) deviations from the azimuthal mean at times $t/t_0 = $ 200, 400, 600, 800 and 1000. \textit{Bottom panels -} Same as above but for Model A. For both models, perturbations become of order $\Delta \rho/ \rho \sim 0.1$ at $t = 400 t_0$. In Model S these are advected out of the simulation domain and the flow reaches steady state around $t = 1000 t_0$. For Model A the density perturbations saturate to $\Delta \rho/ \rho \sim 1$ and $\Delta v / v \sim 0.5$}
    \label{fig:2D_perturbations}
\end{figure*}

We characterize the azimuthal symmetry breaking with the parameter
\begin{equation}
    \Delta X = \frac{X - \langle{X\rangle}_{\theta}}{\langle{X\rangle}_{\theta}},
\end{equation}
where $\langle{X\rangle}_{\theta}$ is the azimuthal mean. In Fig. \ref{fig:2D_perturbations} we plot $\Delta \rho$ (left panels) and $\Delta v$ (right panels) for Model S (top panels) and Model A (bottom panels). We choose a series of times during which the clumpiness is increasing, $200 < t/t_0 < 1000$. The density perturbations are $|\Delta \rho|\lesssim 1$ whereas the velocity deviations are smaller with $|\Delta v|\lesssim 0.5$.  For $t > 1000 \ t_0$, Model S reaches a steady, spherically symmetric state whereas Model A is variable but with a statistically fixed clump distribution.

\begin{figure}
    \centering
    \includegraphics[scale=0.4]{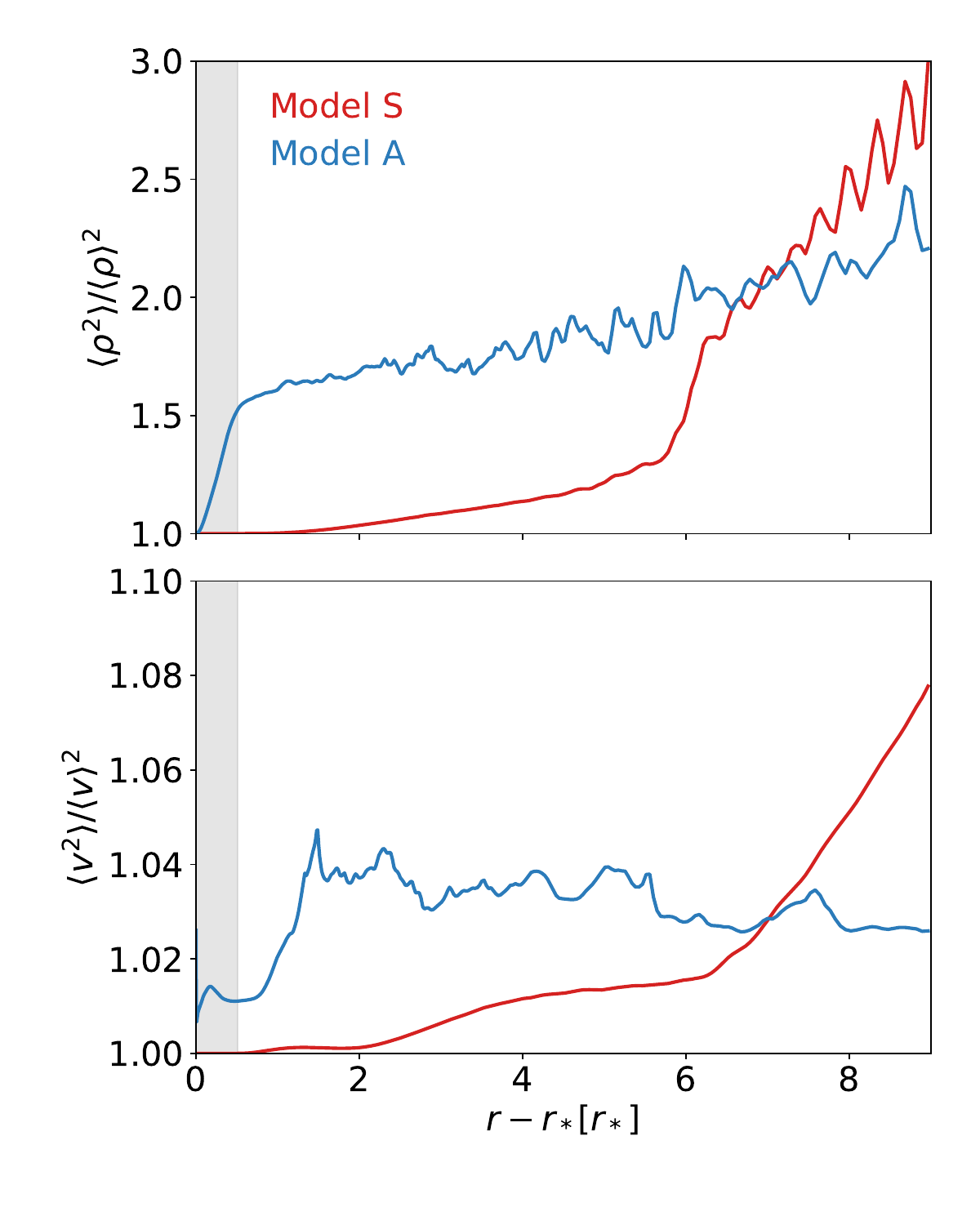}
    \caption{\textit{Top panel -}Time-averaged clumpiness parameter in density (top panel) and velocity (bottom panel) for Model S (red lines) and Model A (blue lines) for $200 \leq t/t_0 \leq 800$. The grey shading indicates the subcritical region for Model S using the radial streaming approximation. In Model S, clumpiness forms outside the subcritical region, consistent with the clumps being advected away at late times and the flow settling into a steady state. By contrast, Model A undergoes rapid growth in the subcritical region, before settling into a smaller growth regime at larger radii where adiabatic stretching becomes important.}
    \label{fig:clumpiness}
\end{figure}

We quantify the degree of inhomogenity in space and time with the clumpiness parameter
\begin{equation}
f_{X} = \frac{ \langle{X^2\rangle}}{\langle{X\rangle}^2},
\end{equation}
where the average $\langle{X\rangle}$ is taken over $\theta$ and either time or radius, depending on the context. In a uniform, stationary flow we have $f_{X} = 1$. In Fig. \ref{fig:clumpiness} we plot the density (top panel) and velocity (bottom panel) clumpiness parameters, $f_{\rho}$ and $f_{v}$ respectively as a function of radius for both models for $200 \leq t/t_0 \leq 800$. Model S has the largest $f_{\rho}$ and $f_{v}$ in the outer parts of the flow where clumps have advected out. At late times, the clumpiness converges to $f_{X} \rightarrow 1$ as the wind reaches a steady state. Model A has the largest increase in density clumpiness in the sub-critical part of the flow. It stays relatively flat throughout the rest of the wind.   

\begin{figure}
    \centering
    \includegraphics[scale=0.4]{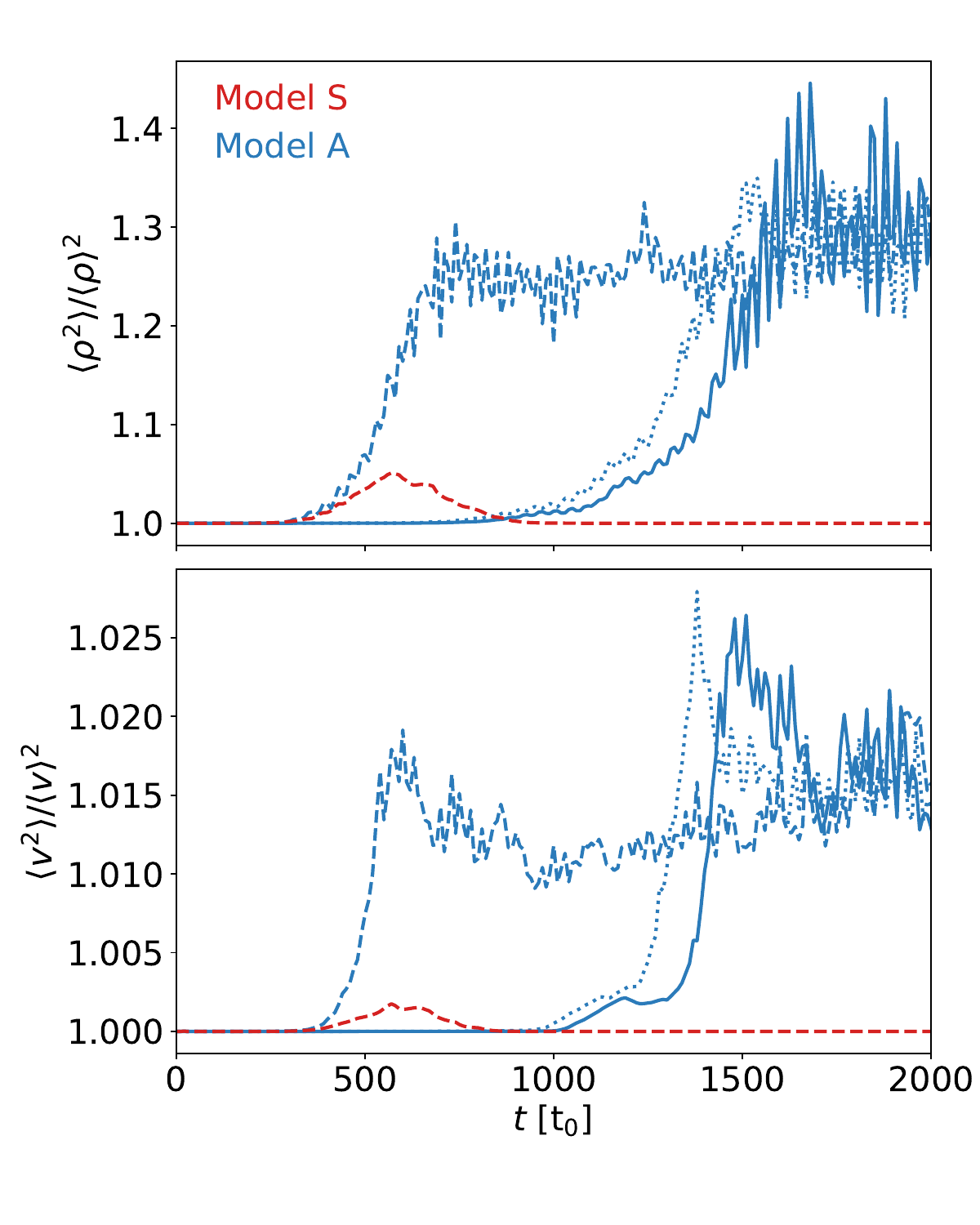}
    \caption{Growth of density (top panel) and velocity (bottom panel) clumpiness as a function of time for Model S (red) and Model A (blue) for initial perturbations $\Delta \rho / \rho = 0$ (solid), $10^{-2}$ (dashed) and $10^{-4}$ (dash-dot). Model S has perturbations growing before being advected out of the domain and converging to a stationary solution. Model A has perturbations grow at a rate independent of initial perturbation size and saturate to a constant value.}
    \label{fig:growth_rate}
\end{figure}

In Fig \ref{fig:growth_rate} we track the clumpiness of the full simulation domain as a function of time. For Model S, we see that the clumpiness peaks at $t \approx 700 \ t_0$, before converging to $f_X = 1$ as the solution settles to a steady state. For Model A, the clumpiness grows exponentially before turning over and saturating to a fixed value. Each curve represents a different simulation with an initial density perturbation amplitude $\Delta \rho /\rho = 10^{-2}$, $10^{-4}$ and $0$. The smaller the perturbation, the longer it takes for $f_{X}$ to saturate. However, we find that regardless of initial pertubation size, the density clumpiness saturates to $f_{\rho} \rightarrow 1.3$ and $f_{v} \rightarrow 1.02$. Further, the growth rate is also independent of the initial size of the perturbation with $df_{\rho}/dt \approx 1.7\times 10^{-4}$ and $df_{v}/dt \approx 1.\times 10^{-3}$.

\begin{figure}
    \centering
    \includegraphics[scale=0.4]{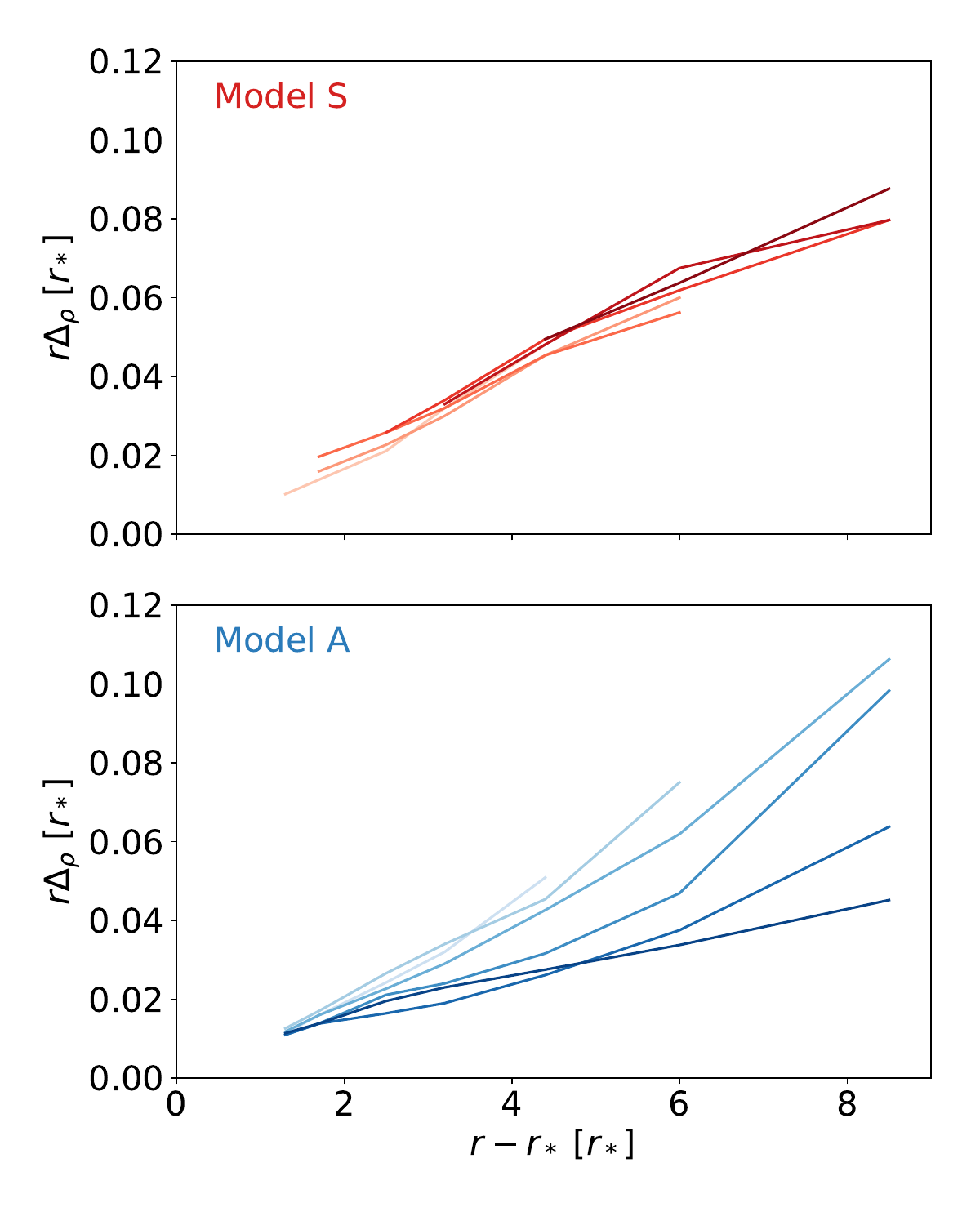}
    \caption{Correlation length of clumps as a function of position in the wind. We plot the azimuthal size $r \Delta \theta_c$ of density features as a function of radial position. Each line indicates a different epoch lasting $\Delta t = 100 \ t_0$, with darker shades indicating later times. For Model S, clump size $\propto r$ as they simply are advected out in the flow. For Model A, this same phenomenon occurs at fixed time. However, clumps also shrink in size, as evidenced by the decrease in slope. This occurs until the clump size reaches approximately the angular resolution.}
    \label{fig:correlation_length}
\end{figure}

To further characterize the clumps we consider the density correlation function
\begin{equation}
    f_c(\Delta) =  \frac{\sum_{t} \sum_{j} \Big(\rho_{j,t} - \langle{ \rho \rangle} \Big) \Big( \rho_{j-\Delta,t} - \langle{ \rho \rangle}\Big)}{\sum_{t} \sum_{j} \Big( \rho_{j,t} - \langle{ \rho \rangle} \Big)^2}, 
    \label{eq:correlation_function}
\end{equation}
where the first index $j$ is the azimuthal position and the second $t$ is time. The average $\langle \rangle$ is over azimuth and time interval $\Delta t = 100 \ t_0$. Using (\ref{eq:correlation_function}) we compute the density correlation function at various radii in the flow and different epochs. For positions and times with a well defined peak, we fit it to a Gaussian profile centered on $\Delta = 0$ and extract the full width at half maximum (FWHM) $\Delta_{\rho}$ and use this as a proxy for the angular size of the clump. The physical clump size is then $r \Delta_{\rho}$, with $r$ the radial position of the clump. In Fig \ref{fig:correlation_length} we plot the physical clump size as a function of position for times $200 \leq t / t_0 \leq 800$ for Model S (top panel) and Model A (bottom panel). Each color shade represents a different epoch lasting $\Delta t = 100 t_0$ and later times shown with darker lines. For Model S, clump size $\propto r$ as they simply are advected out in the flow. For Model A, this same phenomenon occurs at fixed time. However, clumps also shrink in size, as evidenced by the decrease in slope. This occurs until the clump size reaches approximately the angular resolution. Our physical model for the radiation transfer and line force breaks down at the Sobolev scale. The results for Model A therefore suggest that to resolve the late time clump distribution will require accurate modeling on sub-Sobolev scales, including radiation transfer, as clumps tend to break up to this length scale.

\subsection{Source of Instability}

\begin{figure}
    \centering
    \includegraphics[scale=0.42]{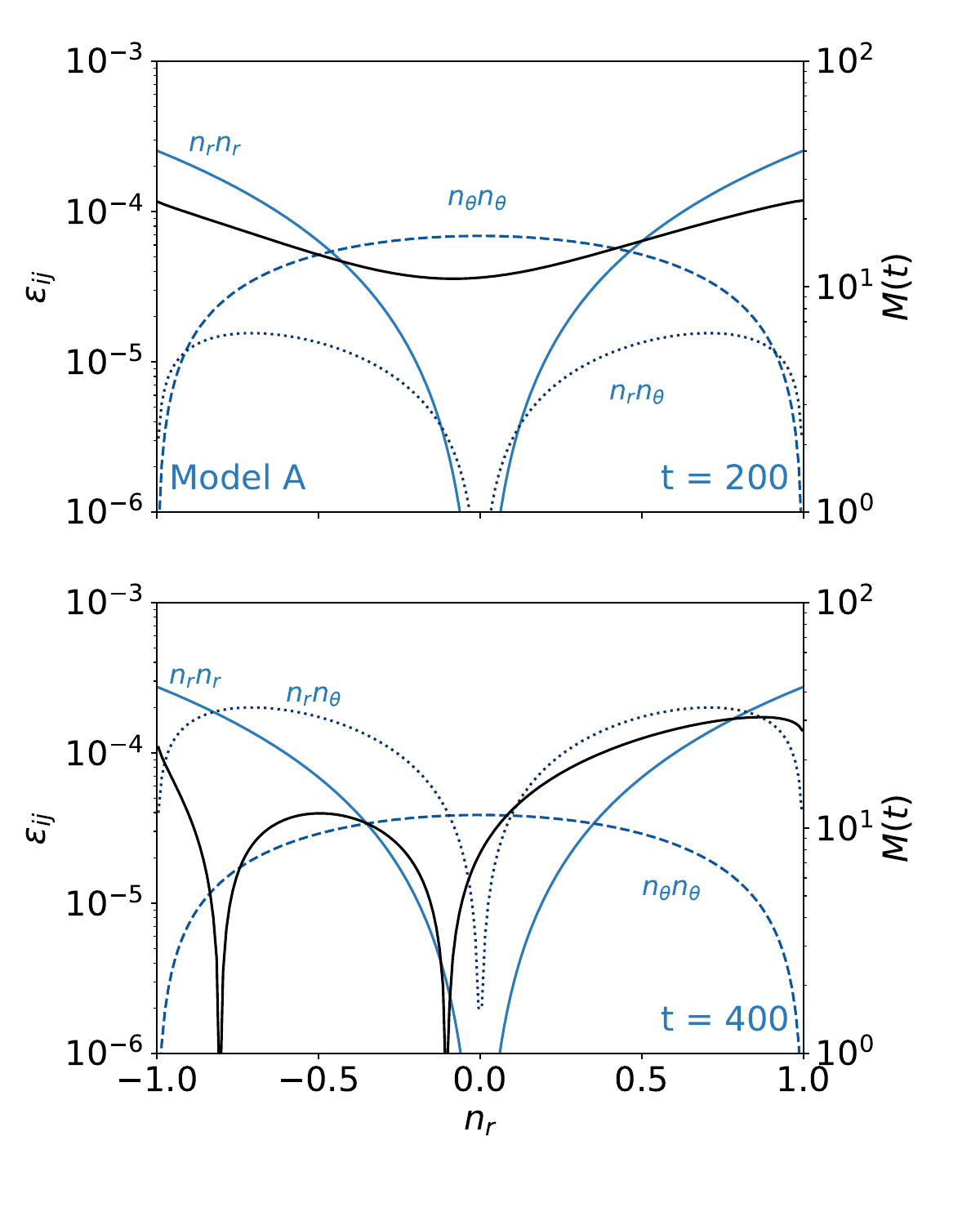}
    \caption{Components of the shear tensor used to compute the velocity gradient in Eq. (\ref{eq:epsilon_ij}) for Model A at a representative point in the sub-critical part of the flow (blue lines) and force multiplier contribution from that direction (solid black line). At early times, $t = 200 \ t_0$ the $rr$ component (solid blue line) dominates for $|n_r| \lesssim 1$ and the $\theta \theta$ component (dashed blue line) dominates for $|n_r| \gtrsim 0$. This results in a smooth force multiplier in all directions. At later times, $t = 400 \ t_0$, there is a change in the hierarchy of scales and the $r \theta$ component (dotted blue line) becomes comparable to the $rr$ or $\theta \theta$ in the appropriate directions. This allows for cancellations in the velocity gradient and a corresponding drop in the force multiplier.}
    \label{fig:epsilon_ij}
\end{figure}

Crucially Model S perturbations form exterior to the critical point. This allows the flow to readjust and advect them away. In Model A the perturbations continue to be generated interior to the critical point. New perturbations form in the sub-critical part of the flow before advecting out of the domain. This can be seen in the density plots (see Fig. \ref{fig:perturbations}) where there is an absence (presence) of perturbations in the shaded grey region.

To further identify the source of these perturbations in the sub-critical pat of the flow we tested a model where the optical depth parameter was calculated in the radial streaming approximation (\ref{eq:dvdr}). With this approximation, Model A reaches a steady state, indicated by the dashed black line in Fig \ref{fig:perturbations}. This suggests that the growth of fluctuations is due to the non-rr components of the shear tensor in Eq (\ref{eq:epsilon_ij}).

We choose a point in the sub-critical part of the flow, at $r = 1.16 r_*$ and $\theta = \pi/2$. Restricting ourselves to 2D, we may express the normal vector $(n_r,n_{\theta}) = (n_r,\sqrt{1. - n_r^2})$. In Fig \ref{fig:epsilon_ij} we plot the relative contributions of each term $\epsilon_{ij}$ in equation (\ref{eq:dvdl}) as a function of $n_r$. At early times, $t = 200 \ t_0$ (top panel), both models have the same hierarchy of scales, with $\epsilon_{\theta \theta}$ dominant for $|n_r| \gtrsim 0$ and $\epsilon_{r r}$ dominant for $|n_r| \lesssim 1$ (only Model A is shown for clarity). For Model S, this hierarchy is maintained throughout the simulation inside the sub-critical part of the flow. However, at $t = 400 \ t_0$ (bottom panel) Model A sees a change in hierarchy whereby the $\theta r$ component becomes comparable to the $rr$ component for  $|n_r| \lesssim 1$. Likewise the   $\theta r$ component is comparable to $\theta \theta$ for $|n_r| \gtrsim 0$. This allows for a cancellation of terms in the velocity gradient (Eq. (\ref{eq:epsilon_ij})) and a drop in the force multiplier $M(t) \propto |dv/dl|^{\alpha}$. Further, we see the $\theta r$ component is dominated by the azimuthal gradients in the radial velocity 
\begin{equation}
    \epsilon_{\theta r} \approx \frac{1}{r} \frac{dv_r}{d\theta},
\end{equation}
i.e velocity gradients along different radial lines of flow. The vanishing velocity gradient manifests in a very small force multiplier and a drop in the \emph{radial} gas acceleration for the line force for rays coming from a certain direction. This decreased radial force serves to further shear the gas in the $\theta$ direction and the velocity gradient with neighbouring radial flows grows further.

Model S experiences a similar behaviour in the super-critical pat of the flow at times when density perturbations are present. The sub-critical part of the flow however maintains the hierarchy of scales seen at early times in Model A, as shown in the top panel of Fig \ref{fig:epsilon_ij}.

\section{Discussion}
\label{sec:discussion}

In the current work we have restricted ourselves to studying continuum radiation transport and treated the line force in the Sobolev approximation. The radiation force due to lines can then be computed from \emph{local} quantities i.e the local velocity gradients and density, rather than more complex treatments that involve radiation transport of the line photons through the entire flow. This has allowed us to confirm many results of instabilities in line driven outflows predicted from linear perturbation calculations.

\cite{Owocki1984} showed that in the absorption regime, the amplification time for radial perturbations is short compared to the dynamical time so perturbations can grow in the acceleration region of the flow. A follow up study \cite{Owocki1985} extended this analysis to include scattering effects and showed that a ``line drag" effect reduces this instability in the base of the wind but not in the outer parts of the flow. Scattering effects can propagate wind disturbances but not make the wind unstable.

These predictions are consistent with our findings. In Model A we find perturbations growing in the base of the wind. Once beyond the critical point, the super-Sobolev perturbations are advected out, but do not grow significantly. In Model S, we do not see growth of perturbations in the sub-critical part of the flow, but initial perturbations are propagated outwards, before the flow settles into a steady state. We carried out an additional simulation with both scattering and absorption effects ($\kappa_s = \kappa_E = \kappa_F = \kappa_{es}$) and found that the resulting wind did not reach a steady state, akin to Model A. Thus, unlike the linear stability analysis we did not find that scattering had a strong stabilizing effect.

The linear perturbation analysis was extended in \cite{Rybicki1990} to include finite disc effects and scattering effects from lateral velocity components. They found that lateral waves are unstable but less so than radial waves. We have found that instabilities in Model A are generated by lateral photons, but that such perturbations are mostly in the radial direction. In other words, the $dv_r/d \theta$ is responsible for decreases in the \emph{radial} component of the radiation force. 

Though azimuthal gradient effects are crucial in generating clumps in the radial directions, the azimuthal velocity remains small and the flow to excellent approximation purely radial. \cite{Dessart2004} explored the effects of a local azimuthal component to the radiation force in the context of co-rotating interaction regions (rotating hot spots on the star) and likewise found that resulting azimuthal velocities were small though the dynamics was strongly affected by the rotation of the star in their case. 

We find the correlation length of density perturbations in Model A becomes smaller, reaching $\sim l_{sob}$ at late times. This suggests that our treatment of the line transfer is breaking down at late times. Clumps, despite being generated on super-Sobolev scales tend to fragment to the Sobolev length. Therefore, correctly resolving the clumpy nature of these flows inevitably requires resolving both the hydrodynamics and radiation transfer at sub-Sobolev scales. 

We ran additional simulations identical to Model A but with $\kappa_E = \kappa_F =  0.1, 0.2, 0.3, 0.4 $ and $0.5 \ \kappa_{es}$. For $\kappa_E = \kappa_F \gtrsim 0.4 \ \kappa_{es}$ we find that the instability persists. However, for $\kappa_E = \kappa_F  \lesssim 0.3 \ \kappa_{es}$ the flow reaches a steady state. As in Model S, perturbations are advected out of the domain and the flow settles to the CAK solution.  Interestingly, the case $\kappa_E = \kappa_F = 0.2 \ \kappa_{es}$ reaches a steady, non-spherical solution where density fluctuations $\Delta \rho/ \rho \lesssim 0.5$ persist in the flow.

Likewise, the instability turns off if the ratio between the flux mean and energy mean opacity changes. With $\kappa_E/\kappa_F = 0.5$ the instability persists. But lowering it down to $\kappa_E/\kappa_F = 0.2$ perturbations advect out of the domain and we reach a steady state. This is expected, since when $\kappa_E/\kappa_F = 0$ we recover the scattering limit of Model S.   

We have explored the parameter space of opacities primarily in the interest of understanding how instabilities can form in outflows driven by radiation pressure. In a physical plasma, we expect the opacity to be dominated by resonant scattering, $\kappa_s$. The contribution to $\kappa_E$ and $\kappa_F$ in the continuum will primarily be due to bound-free processes which will be subdominant. As such, values of $\kappa_E, \kappa_F \gtrsim 0.4 \ \kappa_{es}$ or $\kappa_F \gg \kappa_E$ where we have found this instability to operate may be outside the physical parameter space of most astrophysical plasmas. Determining realistic values for these parameters will require more complete photoionization modeling.

Recent Monte Carlo simulations by Higginbottom et al (submitted to MNRAS) found that a 2D, spherically symmetric OB star wind formed azimuthal density structures in the flow. Their radiation transfer code treats lines as delta functions, so the density structures cannot be forming due to the LDI. Their findings, using a different radiation transfer method, suggest that density features may form due to continuum radiation transfer. 

Studying radiation driving in stellar winds and their associated clumpiness requires modeling radiative transfer at a range of length scales. Firstly, continuum radiation most be resolved on the length scales of the entire flow - the continuum spectra will affect the ionization structure and as well as the momentum available to transfer to the outflowing gas. Secondly, line transfer at or below the Sobolev length will determine how strongly the radiation field couples to the gas. We cannot trust our photoioniation modeling if we get the continuum radiation transfer wrong and likewise we cannot properly account for any physics at or below the Sobolev length, where clumpiness is thought to occur, if we do not properly resolve the line transfer. In this work we attack the problem of studying instabilities that occur due to continuum radiation transfer \emph{above} the Sobolev length scale. In a future work, this will allow us to study radiation transfer below the Sobolev length and disentangle the effects of continuum and line radiation transfer. We will model the radiation transfer of a single line and study its effects on generating perturbations in the wind. The results from this study suggest that treating the continuum radiation transfer solely due to scattering opacities will not grow perturbations in the base of the wind. Thus, we may better isolate the effects of a single absorption line on the generation of clumps in the sub-critical parts of the flow.

\section*{Acknowledgments} 
Support for this work was provided by the National Aeronautics and Space Administration under TCAN grant 80NSSC21K0496. We thank Tim Kallman, Daniel Proga, Yan-Fei Jiang and the entire DAWN TCAN collaboration for fruitful discussions.

\section*{Data Availability Statement}
The simulations were performed with the publicly available code \textsc{Athena++} available at https://github.com/PrincetonUniversity/athena The authors will provide any additional problem generators and input files upon request.

%%%%%%%%%%%%%%%%%%%%%%%%%%%%%%%%%%%%%%%%%%%

\bibliographystyle{mnras}
\bibliography{progalab-shared}

\bsp	% typesetting comment
\label{lastpage}

\end{document}